\documentclass[pre,reprint,amsmath,amssymb, aps, superscriptaddress]{revtex4-1}
\usepackage[utf8]{inputenc}
\usepackage{balance}
\usepackage[T1]{fontenc}
\usepackage{newtxtext}
\usepackage[varvw]{newtxmath}
\usepackage{graphicx}
\usepackage{color}
\usepackage{amsmath,amssymb}
\usepackage{fixmath}
\usepackage{array}
\usepackage{bm}
\usepackage{ulem}

\begin{document}

\title{Nonequilibrium diffusion of active particles bound to a semi-flexible polymer network: simulations and fractional Langevin equation}
\author{Hyeong-Tark Han}
\affiliation{Department of Physics, POSTECH, Pohang 37673, Republic of Korea}
\author{Sungmin Joo}
\affiliation{Department of Physics, POSTECH, Pohang 37673, Republic of Korea}
\author{Takahiro Sakaue}
\affiliation{Department of Physical Sciences, Aoyama Gakuin University, Kanagawa 252-5258, Japan}
\email[]{sakaue@phys.aoyama.ac.jp}
\author{Jae-Hyung Jeon}
\affiliation{Department of Physics, POSTECH, Pohang 37673, Republic of Korea}
\affiliation{Asia Pacific Center for Theoretical Physics, Pohang 37673, Republic of Korea}
\email[]{jeonjh@postech.ac.kr}


\begin{abstract}
In a viscoelastic environment, the diffusion of a particle becomes non-Markovian due to the memory effect. 
An open question is to quantitatively explain how self-propulsion particles with directional memory diffuse in such a medium. 
Based on simulations and analytic theory, we address this issue with active viscoelastic systems where an active particle is connected with multiple semi-flexible filaments.
Our Langevin dynamics simulations show that the active cross-linker displays super- and sub-diffusive athermal motion with a time-dependent anomalous exponent $\alpha$. In such viscoelastic feedback, the active particle always has superdiffusion with $\alpha=3/2$ at times shorter than the self-propulsion time ($\tau_A$). At times greater than $\tau_A$, the subdiffusion emerges with $\alpha$ bounded between $1/2$ and $3/4$. Remarkably, the active subdiffusion is reinforced as the active propulsion (Pe) is more vigorous. In the high-Pe limit, the athermal fluctuation in the stiff filament eventually leads to $\alpha=1/2$, which can be misinterpreted with the thermal Rouse motion in a flexible chain. 
We demonstrate that the motion of active particles cross-linking a network of semi-flexible filaments can be governed by a fractional Langevin equation combined with fractional Gaussian noise and an Ornstein-Uhlenbeck noise.
We analytically derive the velocity autocorrelation function and mean-squared displacement of the model, explaining their scaling relations as well as the prefactors. 
We find that there exist the threshold Pe ($\mathrm{Pe}^*$) and cross-over times ($\tau^*$ and $\tau^\dagger$) above which the active viscoelastic dynamics emerge on the timescales of $\tau^* \lesssim t \lesssim \tau^\dagger$.  
Our study may provide a theoretical insight into various nonequilibrium active dynamics in intracellular viscoelastic environments.
\end{abstract}
\maketitle

\section{Introduction}

Active agents that illustrate nonequilibrium kinetic motion are ubiquitous in the microscopic world, such as molecular motors, active colloids, and unicellular organisms~\cite{Wu2000, 
Marchetti2013, 
Lauga2012, %
Chen2015,Gal2013,Bechinger2016}.
They are often found in living systems or have been experimentally realized using chemical reactions~\cite{Howse2007,Volpe2011,Palacci2010}, magnetic forces~\cite{Dreyfus2005,Tierno2008,Ghosh2009}, diffusiophoresis~\cite{Buttinoni2013,Saad2019}, electrohydrodynamic forces~\cite{Bricard2013}, or droplets in a biocompatible
oil~\cite{Izri2014}.
The active agents typically perform a stochastic movement with a directional persistence on a certain timescale, and their diffusion does not obey the Einstein relation.  
It has been shown that the kinetic motion of these active agents can be successfully described by stochasticity-based mesoscopic models, e.g., run-and-tumble particle~\cite{Kafri2008,Tailleur2008,Matthus2009,Lauga2012}, active Brownian particle~\cite{Howse2007,tenHagen2011,Romanczuk2012,Zheng2013}, or active Ornstein-Uhlenbeck particle (AOUP) models~\cite{Wu2000,Samanta2016,Bechinger2016,Nguyen2021}.
These models have enabled one to quantitatively explain experimentally measurable active dynamics, including mean-squared displacement (MSD), long-time diffusivity, and the probability density function of displacements. 

Beyond the question of an isolated active particle, extensive investigations have been conducted on the theme of active particles in complex environments. For example, experimental and computational studies were devoted to investigating collective behaviors of a collection of active particles~\cite{Budrene1995,Brenner1998,
Cates2015,Omar2021,Alert2022} or the mixtures of active particles and Brownian particles~\cite{Angelani2011,Stenhammar2015,Wysocki2016,Lozano2019}, complex diffusion of active particles in porous media~\cite{Bhattacharjee2019,Creppy2019,Kurzthaler2021,Kjeldbjerg2022}, polymer networks~\cite{Toyota2011,Cao2021,Kim2022,Kumar2023} or polymer solutions \cite{Loi2011,Weber2012,Yuan2019,Du2019}.  
These studies led to the discoveries of unexpected nonequilibrium transport dynamics of active particles and their pattern formation, further giving insights into the complex dynamics of living systems that emerge from the interactions of active agents and other components of the systems~\cite{Bronstein2009,Jeon2011,Weigel2011,Song2018}. 

Among them, active viscoelastic systems may be of great interest 
as they are intimately connected to the intracellular environments or nonequilibrium bio-polymer networks or gels where active agents and biofilaments coexist. Examples include the actin-myosin polymer network~\cite{ Harada1987,Amblard1996,Wong2004,Pollard2009}, molecular motor-driven transports in a cell~\cite{Caspi2000,Weihs2006,Wilhelm2008,Kahana2008,Reverey2015,Chen2015,Song2018}, loci or telomere motion in chromosome DNAs~\cite{Wang2008,Bronstein2009,Bronshtein2015,Stadler2017,Ku2022,Kimura_2022}, the cross-link in endoplasmic reticulum networks~\cite{Vale1988,Lin2014,Speckner2018}, micro-swimmers in a hydrogel~\cite{Lieleg2010,Caldara2012,Bej2022}, etc.    
In these systems, the embedded polymers render the viscoelastic environment, giving rise to complex kinetic motion of the (active) particle bound to the polymer system or diffusing through it due to the long-time memory effect. There have been not a few experimental and computational studies reporting novel active viscoelastic dynamics in various contexts~\cite{Liverpool2001,Humphrey2002,Sevilla2019,Lozano2019}, but the description was phenomenological or based on intuitive arguments. When complex interactions come into play, solving an active model (e.g., run-and-tumble, active Brownian particle, and AOUP) with other interacting components is highly nontrivial. It is yet an open question to establish a mesoscopic, quantitative theory to describe active viscoelastic motion.

Here we study an active viscoelastic system comprising an active particle and multiple semiflexible filaments [Fig.~\ref{fig1}] and also establish a Langevin-type mesoscopic theory for the active motion observed in simulation. Prior to the current study, we theoretically investigated an active viscoelastic system where an active particle is connected to multiple flexible chains~\cite{joo2020}. It turns out that the flexible polymer environment gave nontrivial viscoelastic feedback on the active tracer's movement. The active diffusion of the tracer became a Gaussian anomalous diffusion of the MSD scaling as $\langle \Delta x^2(t)\rangle\propto t^\alpha$ where the anomalous exponent $\alpha$ gets monotonically smaller as the self-propulsion speed is larger. 
We analytically solved the coupled Langevin equation for an active particle cross-linking the Rouse chain, finding that the motion of the active particle is the superposition of a Rouse thermal motion ($\langle \Delta x^2(t)\rangle\propto t^{1/2}$) and an ultra-slow athermal motion ($\langle \Delta x^2(t)\rangle\propto \log t$) where the amplitude of the athermal component gets stronger with the propulsion speed. Importantly, the viscoelastic athermal subdiffusion of the active tracer (cross-linker) bound to a flexible chain can be described by the following fractional Langevin equation:
\begin{equation}\label{FLEflexible}
\int_{-\infty}^t dt'K(t-t')\dot{X}(t')=\Gamma(1/2)\frac{d^{1/2}X}{dt^{1/2}}=\psi_\mathrm{th}(t)+\psi_\mathrm{ac}(t)
\end{equation}
where $X(t)$ is the position of the active particle, $K(t)$ is a power-law decaying memory kernel leading to $\frac{d^{1/2}X}{dt^{1/2}}$, i.e., the Caputo fractional derivative of order $1/2$, and $\psi_\mathrm{th}$ and $\psi_\mathrm{ac}$ are respectively the thermal and active noises in this coarse-grained level.  

Initiated from this work, we extend our scope into a more complex system where an active particle is bound to (or equivalently cross-links) a semi-flexible polymer network [Fig.~\ref{fig1}]. Our aim is to systematically study the active diffusion in a semi-flexible filament network compared to that in a flexible environment and to examine whether the above Langevin-type formalism can still be developed. Biologically, the active semi-flexible system is very interesting in the sense that many of the bio-polymers are semi-flexible chains, which are known to play a crucial role in the cell~\cite{Harada1987,
marko1995stretching,Ghosh2014,
Bausch2006,Brangwynne2008,Schaller2010}.
We perform the Langevin dynamics simulation for our model. It shows that the active diffusion associated with the semiflexible filament has several features that are clearly distinguished from the flexible case. The diffusion dynamics attains a distinctive regime of super- and sub-diffusion separated by the timescale of the self-propulsion time. Instead of the ultra-slow athermal diffusion in the flexible case, the athermal viscoelastic motion in the semi-flexible filament results in a seemingly Rouse-like dynamics ($\langle \Delta x^2(t)\rangle\propto t^{1/2}$). This means that the Rouse motion in living systems should be carefully interpreted with multiple observables so as to distinguish between genuine thermal motion and a nonequilibrium directional motion bound to a stiff chain.
We demonstrate that a fractional Langevin equation can be formulated using the tension propagation theory \cite{sakaue2017active,Vanderzande_2019} and the generalized Langevin equation formalism~\cite{Corts1985,Panja2010,Jeon2013,Metzler2014,Saito_2015,Vandebroek2015,Wu2018} that describes the stochastic motion of an active particle under the viscoelastic feedback from the semiflexible filament. It turns out that the nonequilibrium fractional Langevin equation has the same structure as Eq.~\eqref{FLEflexible} but with a different form of the memory kernel.





This paper is organized as follows. 
In Sec.~\ref{sec:model}, we introduce our model system where an active particle cross-links multiple semi-flexible filaments and explain the model for active particles and simulation details. 
In Sec.~\ref{sec:result.ABP}, we provide the results of our Langevin dynamics simulation. We investigate dynamic quantities such as MSDs, anomalous exponent, and displacement autocorrelation function for varying the simulation parameters including the self-propulsion speed, persistence length of the filament, the connectivity of the network, and boundary conditions. 
In the following section \ref{sec:theory}, we construct a nonequilibrium fractional Langevin equation as an effective theory. Using this we obtain analytic expressions for MSDs, velocity autocorrelation functions, and displacement autocorrelation functions, which excellently explain the observed simulation results. We also provide an analysis of the time- and propulsion-dependent dynamics of the active cross-linker. 
Finally, in Sec.~\ref{sec:summary}, we summarize the main results with a discussion.

\section{The Model} \label{sec:model}
\begin{figure}[h]
\centering
\includegraphics[width=9cm]{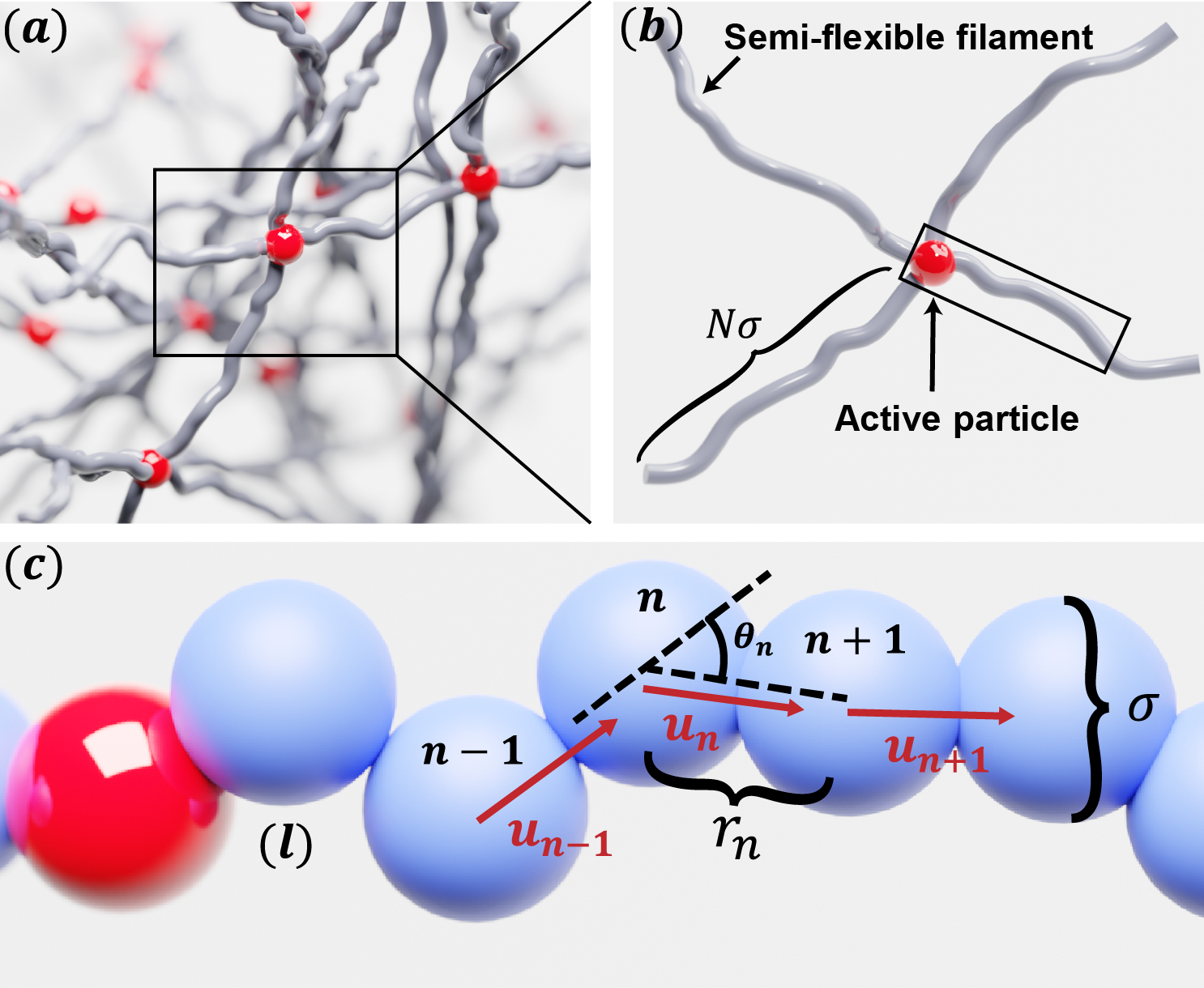}
\caption{The schematics of the active semi-flexible polymer system under consideration. 
(a) A semi-flexible polymer network (e.g., made of actin filaments, microtubules, or chromosomes) together with active particles (red) attached to the network. The active particles serve as a cross-linker connecting $f$ polymer arms ($f$: the functionality). 
(b) The geometry of active star polymer, i.e., a unit motif of the above active polymer network. In this setup, the active particle is the cross-linker connecting $f$ semi-flexible filaments each of which has the length of $N$ segments ($\sigma$: inter-monomer distance). The end segment of each filament is free to move in the case of free-boundary simulation (the phantom network model) or fixed in space in the case of fixed-boundary simulation (the affine network model). 
(c) The zoom-in illustration of the semi-flexible filament. The discrete beads (blue) comprise an inextensible wormlike chain of a bending persistence length $l_p/\sigma$. The $\mathbf{u}_{n}$ is the unit tangent vector of the $n$th monomer and $\theta_n$ is the bending angle between $\mathbf{u}_{n-1}$ and $\mathbf{u}_{n}$. 
}
\label{fig1}
\end{figure}


The active viscoelastic system considered in this work is the following [Fig.~\ref{fig1}]. The active particle is attached to a semiflexible polymer network such that its self-propelled diffusion disturbs the polymer configuration which, in turn, gives rise to long-time negative feedback to the active motion itself. To simulate this system, we constructed the star-like polymer in Fig.~\ref{fig1}(b) where a self-propelling particle is connected to semiflexible filaments as a cross-linker. The present system is the semiflexible polymer version of the (flexible) star polymer investigated in our previous work~\cite{joo2020}. In the simulation, the self-propelled active particle was modeled by the active Ornstein-Uhlenbeck particle (AOUP)~\cite{Wu2000,Samanta2016,Bechinger2016,Nguyen2021}, which is governed by the Langevin equation 
\begin{equation}
\gamma \dot{\mathbf{R}}(t) = \bm{\xi}(t) + \bm{\eta}(t).
\label{eq:abp}
\end{equation}
Here, $\bm{\xi}$ is a delta-correlated thermal force having the covariance $\langle \xi_\mu(t)\xi_\nu(t')\rangle = 2\gamma k_BT\delta_{\mu\nu}\delta(t-t')$ ($\mu$ and $\nu$ are the Cartesian component,  $k_B$ the Boltzmann constant, and $T$ is the absolute temperature). 
$\bm{\eta}$ is an active force leading to the self-propelled motion with the propulsion speed $v_p$ and the self-propulsion memory time $\tau_A$, which satisfies the Ornstein-Uhlenback process
\begin{equation}
    \dot{\eta}_\mu(t)=-\frac{1}{\tau_A}\eta_\mu(t) + \sqrt{\frac{2\gamma^2v_p^2}{d\tau_A}}\zeta(t)
    \label{eq:active-Gaussian noise}
\end{equation}
where $\zeta(t)$ is a delta-correlated Gaussian noise of unit variance. The active force then has the properties of $\langle \eta_\mu(t)\rangle=0$ and $\langle \eta_\mu(t) \eta_\nu(t')\rangle = \frac{1}{d}\gamma^2 v_p^2 \delta_{\mu,\nu}\exp{\left(-\frac{|t-t'|}{\tau_A}\right)}$ [$d$: the dimensionality]. 

The semi-flexible filament is made of an inextensible wormlike chain (WLC)~\cite{marko1995stretching,Sait1967}. The bending energy of an $N$-segmented chain (of the bending persistence length $l_p$) is given by the effective Hamiltonian    
$E_{\mathrm{cv}}=\frac{k_BTl_p}{\sigma}\sum_{n=2}^{N-1} \left(1-\cos{\theta_{n}}\right)=\sum_{n=2}^{N-1}E_{\mathrm{bend}}(\theta_{n})$ where $\theta_n$ is the angle made by the successive tangent vectors [Fig.~\ref{fig1}(c)] and $\sigma$ is the average inter-monomer distance. To implement the inextensible filament, a bond potential between successive segments was introduced using the expanded finite extensible nonlinear elastic (FENE) potential~\cite{warner1972},
$E_{\mathrm{bond}}(r_{n})=-\frac{1}{2}KR_0^2 \ln\left[1-\left(\frac{r_{n}-\sigma}{R_0}\right)^2\right]
$, where $r_{n}=|\mathbf{R}_{n+1}-\mathbf{R}_n|$ is the distance between neighboring monomers, $K$ the spring constant, and $R_0+\sigma$ is the maximum extent of the bond.
 
Using the AOUP and the semi-flexible filament, we constructed an AOUP-WLC composite system where the central AOUP cross-links the $f$ arms of $N$-segmented wormlike filaments [Fig.~\ref{fig1}(b)]. Because the star-polymer mimics the repeat motif of a polymer network, we suppose that $f/2$ semiflexible filaments are cross-linked at the center at which the bending energy occurs only within the same filament.  
Under this consideration, the total Hamiltonian for the AOUP polymer composite is written as 
\begin{equation}
E_{\mathrm{tot}}= \sum_{l=1}^{f/2}\sum_{n=-(N-1)}^{N-1}E_{\mathrm{bend}}^{(l)}(\theta_{n})+\sum_{l=1}^{f/2}\sum_{n=-N}^{N-1}E_{\mathrm{bond}}^{(l)}(r_{n})
\label{eq:total-star}
\end{equation}
where $n\in\{-N,\ldots,N\}$ and $n=0$ is the AOUP cross-linker. With $l_p\ll L(=N\sigma)$ the polymer can be considered as a flexible chain, which then reproduces the flexible star polymer system we previously studied~\cite{joo2020}. In the limit of $l_p/\sigma \gg 1$, the semi-flexible polymer becomes a rod-like stiff filament. In our model, the filament has a length of $N=50$ and $f=2,~4,~6$, and $8$.

We performed the Langevin dynamics simulation on the AOUP viscoelastic system at various conditions. 
Denoting $\mathbf{R}^{(l)}_{n}=(X_n^{(l)},Y_n^{(l)},Z_n^{(l)})$ as the Cartesian coordinates of the $n$-th monomer in the $l$-th filament [$\mathbf{R}_0=\mathbf{R}_A$ is the AOUP cross-linker], its diffusion dynamics is described by the following overdamped Langevin equation
\begin{equation}
\gamma \dot{\mathbf{R}}^{(l)}_{n}=-\nabla_n^{(l)}E_{\mathrm{tot}} +\bm{\xi}_n(t)+\bm{\eta}(t)\delta_{n,0}.
\label{eq:eom}
\end{equation}
For simplicity, the size of the AOUP was set to be the same as that of the monomers in the filament. We numerically solved the above Langevin equation by the 2nd order Runge-Kutta (Heun) method~\cite{joo2020}. 
In our simulation study, we investigated both cases of the free and fixed boundary conditions (B.C.s) for the end segments of the filament. The free B.C. can be understood as the phantom network model where the end monomers are free to move. The other B.C is the pinned end monomer ($\dot{\mathbf{R}}^{(l)}_{\pm N}=0$), corresponding to the affine network model. It turns out that B.C. barely changed the diffusion dynamics of the AOUP cross-linker in the time scale we are interested in.
Therefore, in this work, we focus on presenting the case of free B.C. if, otherwise, indicated. 

The Langevin simulation was carried out for AOUPs with the P\'eclet number $\textrm{Pe} = \frac{v_p \sigma}{D}$ ranging from $0$ to $80$. The flexibility of the filament varied from $l_p/\sigma=0.5$ to $25$. In the simulation, the basic units are the monomer diameter $\sigma$, the time $\tau=\sigma^2/D$, and the energy $\epsilon=k_B{T}$ (${T}=300$~K). The thermal diffusivity of the polymer bead and AOUP is then $D=\sigma^2/\tau$. 
The integration time is $\Delta t=10^{-3}~\tau$ while the propulsion memory time is set to be $\tau_A=0.1~\tau$. The system was initially equilibriated for $t=10^5~\tau$ to remove the effect of the initial condition [Fig.~\ref{figa1}(a)]. For evaluating physical quantities, we typically performed 100 independent runs with $T=10^5~\tau$ for a given parameter set. 


\section{Simulation: Active diffusion of the AOUP cross-linker} \label{sec:result.ABP}

\begin{figure*}
\centering
\includegraphics[width=15cm]{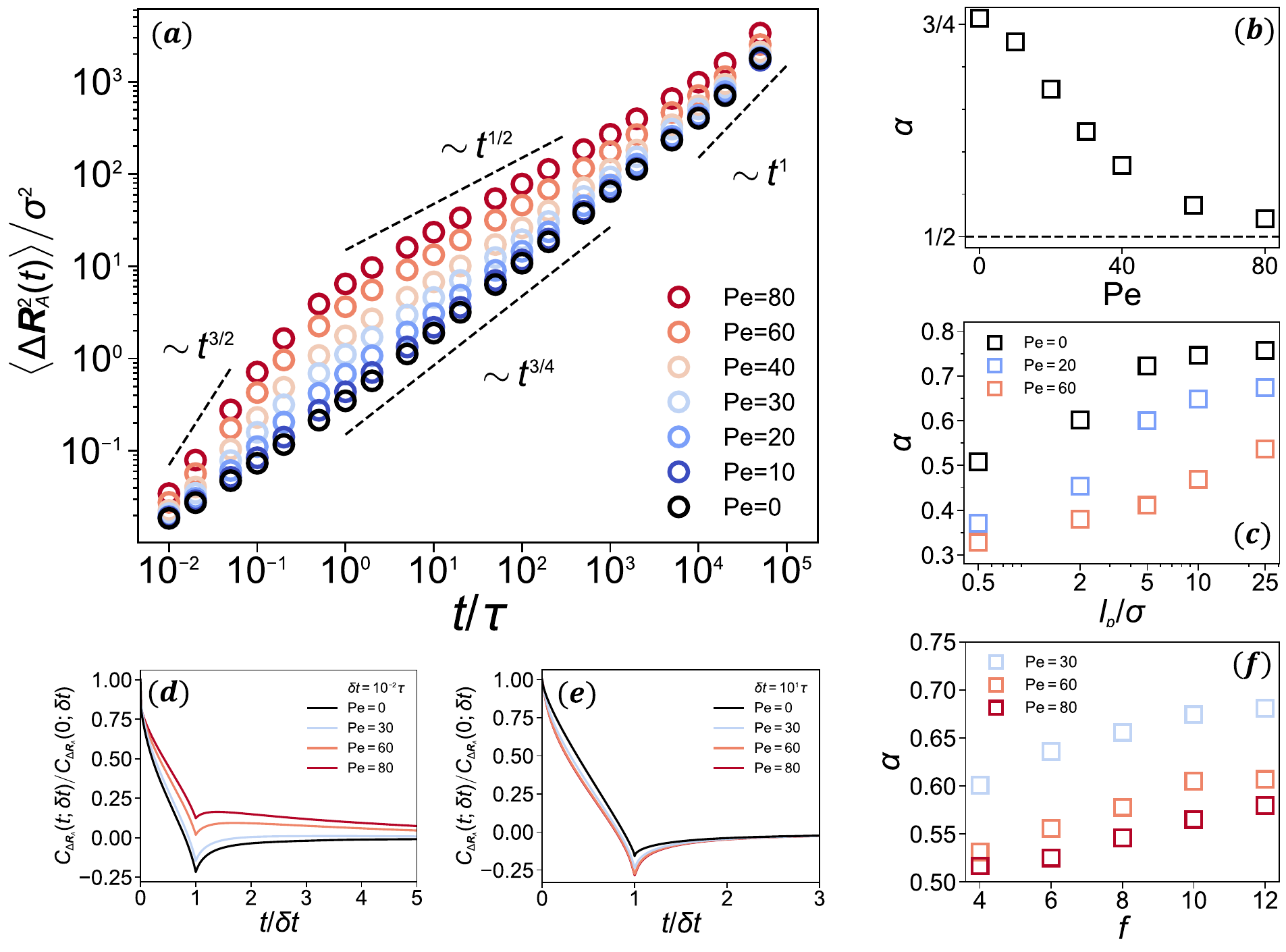}
\caption{Active diffusion of the AOUP cross-linker (the persistence length of the filament: $l_p/\sigma=25$, the functionality $f=4$, and the free boundary condition).   
(a) The MSD curves for the AOUPs at various  $\mathrm{Pe}$ conditions. The dashed lines are the guided scalings explained in the text. The MSD was evaluated from 100 trajectories in the sense of time and ensemble averaging, i.e., $\langle \Delta \mathbf{R}^2_A(t)\rangle\equiv \overline{\langle  [\mathbf{R}_A(t_0+t)-\mathbf{R}_A(t_0)]^2\rangle}_\mathrm{sp}$ where $\overline{\cdots}  \equiv\int_0^{t_M-t}\cdots dt'/(t_M-t)$ represents  the time-averaged MSD from a single particle trajectory and $\langle\cdots\rangle_\mathrm{sp}$ is the ensemble-averaging  over samples~\cite{Metzler2014}.  
(b) The anomalous exponent $\alpha$ as a function of $\mathrm{Pe}$. The $\alpha$s were measured by fitting to the MSD curves in (a) in the time domain $t\in[10^0,~10^2]$ with the fit function $f(t)=A t^\alpha$. (c) The anomalous exponent $\alpha$ as a function of $l_p$. 
(d, e) Displacement autocorrelation function $C_{\Delta \mathbf{R}_A}(t;\delta t)/C_{\Delta \mathbf{R}_A}(0;\delta t)$ of the AOUP for $\mathrm{Pe}=0,~30,~60$, and 80. The plotted lines are the simulation results for the corresponding Pe values. Displacement was $\Delta\mathbf{R}_A(t;\delta t)=\mathbf{R}_A(t+\delta t)-\mathbf{R}_A(t)$ for $\delta t=10^{-2}~\tau$ (d) and $\delta t=10^1~\tau$ (e). 
(f) The variation of $\alpha$ vs the functionality $f$ for the AOUPs ($\mathrm{Pe}=30,~60$, and $80$ and $l_p/\sigma=25$). 
}
\label{fig2}
\end{figure*}

We have simulated the AOUP star polymer with four arms ($f=4$) for varying Pe and $l_p$. 
Fig.~\ref{fig2}(a) shows the simulated MSDs of the AOUP cross-linker in the semiflexible filaments of $l_p=25~\sigma$ for increasing Pe. 
The Brownian limit cross-linker (Pe=0) displays the well-known subdiffusive undulation of a stiff filament with the anomalous exponent $\alpha=3/4$, which is the collective motion of a semiflexible chain occurring on the timescale from $\tau_0$ (the microscopic time that an individual monomer interacts with the neighbor monomers) to $\tau_R$ (the relaxation time). Shorter than $\tau_0$, it does not have a well-defined power-law scaling. In the simulation, empirically, the MSD seems to grow like $t^{0.5}$. Beyond $\tau_R$ the Brownian cross-linker shows Fickian dynamics for the drift of the total system where $\langle \Delta \mathbf{R}^2_A(t)\rangle=6D_Gt$ and $D_G=k_BT/\gamma N_{\mathrm{tot}}$. We define the relaxation time $\tau_R$ such that the subdiffusive monomer dynamics cross-overs to the Fickian dynamics.  The relaxation time then can be found via the equation  $C(\frac{l_p}{\sigma})^{-1/4}(\frac{\tau_R}{\tau})^{3/4}=6(\frac{D_G\tau}{\sigma^2})(\frac{\tau_R}{\tau})$ with a constant $C\approx0.769$ numerically found from the simulation data~\cite{semiPRL2002}. Note that the relaxation time scales as $\tau_R\propto \frac{\sigma^3 N_\mathrm{tot}^4}{D l_p}$. For given star polymer system ($l_p=25\sigma$, $f=4$ and $N=50$), we obtained $\tau_R/\tau \approx 1.6\times 10^4$.  Solving this equation, we obtained $\tau_R/\tau \approx 1.6\times 10^4$ for our polymer network system ($f=4$ and $N=50$). 

The AOUP exhibits the active diffusion clearly deviated from that of the Brownian particle. The MSD has two distinct scalings depending on the timescale as Pe is increased. For $t\lesssim \tau_A$, the AOUP evidently has the superdiffusive motion with anomalous exponent $\alpha\approx 3/2$. For $\tau_A \lesssim t\lesssim \tau_R$, the AOUP has a subdiffusive motion where the $\alpha$ is smaller than the stiff chain's exponent $3/4$. Importantly, the exponent eventually reaches the limiting value of $\alpha\approx1/2$ when Pe becomes sufficiently large [Fig.~\ref{fig2}(b)].  For $t>\tau_R$, the AOUP has a Fickian diffusion, which is attributed to the drift of the center-of-mass of the total system [Fig.~\ref{figa1}(b)]. In this regime, the MSD is given by $\langle \Delta \mathbf{R}^2_A(t)\rangle\sim(6D_G+2v_p^2\tau_A/N_{\mathrm{tot}}^2)t$ where the second term in the parenthesis explains the contribution from the active noise. 

We note that the AOUP cross-linker in  semi-flexible filaments  is qualitatively very different from the counterpart in a flexible polymer that we investigated in the previous work~\cite{joo2020}. In the flexible polymer system, the AOUP has a subdiffusion with  $\alpha\leq1/2$ ($\alpha=1/2$: the Rouse exponent when $\mathrm{Pe}=0$), and $\alpha$ monotonically decreases with increasing Pe. It was found that the active diffusion of the AOUP interacting with a Rouse polymer has the MSD with $\langle \Delta \mathbf{R}^2_A(t)\rangle\sim B(\mathrm{Pe})t^{1/2}+\ln t$ where the $\ln t$ term explains the athermal viscoelastic subdiffusion via the harmonic interactions against the self-propelled motion~\cite{joo2020}. The factor $B\to 0$ as Pe goes to infinity. 
Accordingly, the logarithmic part becomes stronger as Pe is higher and the $\alpha$ for the empirical power-law scaling $\langle \Delta \mathbf{R}^2_A(t)\rangle\propto t^\alpha$ monotonically decreases with increasing Pe from the Rouse exponent ($\alpha=1/2$). In our current semi-flexible filament model, we recovered these viscoelastic active motions in the flexible chain limit when $l_p/\sigma=0.5$ (see the Appendix~\ref{sec:flex} \& Fig.~\ref{figa1}(c)).
In the case of a semi-flexible filament model, the active subdiffusion of the AOUP varies with an anomalous exponent of  $1/2\leq\alpha\leq3/4$. Fig.~\ref{fig2}(b) depicts the variation of $\alpha$ (that fitted from the MSD for $t$ in $[10^0,~10^2]$) as a function of Pe when $l_p/\sigma=25$. It is interesting to note that the semiflexible chain exhibits the seemingly Rouse undulation motion with the MSD of $t^{1/2}$ when the semiflexible filament is strongly driven by the AOUP. 
In experiments, the active undulation dynamics in a semiflexible filament could be misunderstood as the thermal Rouse motion in a flexible polymer network if one does not carefully analyze the data. 
See also Ref.~\onlinecite{Vanderzande_2019}, where the active MSD of $t^{1/2}$ is also predicted for chromatin loci which are driven by active force dipoles.
The observed active subdiffusion is insensitive to the boundary condition. This is confirmed by our supplementary simulation with the fixed boundary condition [Appendix \ref{sec:boundary} \& Fig.~\ref{figa1}(d)]. The two MSDs for the free and fixed boundary conditions are compared. Except for the large times at $t>\tau_R$ (where the boundary effect emerges), the AOUP dynamics in both systems are identical.

We measured the effect of the bending rigidity ($l_p$) on the AOUP dynamics. In Fig.~\ref{fig2}(c) we estimated the $\alpha$ as a function of $l_p$ for the AOUPs at $\mathrm{Pe}=0,~20$, and $60$. In the Brownian limit, $\alpha$ increases from $1/2$ to $3/4$ as the filament stiffness is increased from a flexible chain to a stiff one. In the presence of sufficient active forces ($\mathrm{Pe}\gg 1$), the $\alpha$s are always smaller than the ones at $\mathrm{Pe}=0$; the AOUP in the semiflexible filament appears more subdiffusive than the  thermal motion of the filament. Increasing $l_p$ tends to increase $\alpha$. It is saturated to $\alpha\approx 1/2$ in the stiff filament as Pe is sufficiently high.  

In Fig.~\ref{fig2}(d) \& (e), we studied the viscoelastic correlation effect in the AOUP diffusion. For this, we define the displacement autocorrelation function (DACF), 
\begin{eqnarray}
C_{\Delta \mathbf{R}_A}(t;\delta t)
&\equiv&
\left\langle \Delta \mathbf{R}_A(t;\delta t)  \cdot\Delta \mathbf{R}_A(0;\delta t) \right\rangle   \\ &=&\int_0^{\delta t}dt_1\int_t^{t+\delta t} dt_2  \langle \mathbf{v}(t_1)\cdot \mathbf{v}(t_2) \rangle \label{dacf}
\end{eqnarray}
where $\Delta \mathbf{R}_A(t;\delta t) \equiv \mathbf{R}_A(t+\delta t)-\mathbf{R}_A(t)$ is the displacement at time $t$ during lag time $\delta t$.
The DACF has the same correlation structure as the velocity autocorrelation via Eq.~\eqref{dacf}. 
Positive (negative) correlation in DACF indicates that the AOUP performs a persistent (anti-persistent) random walk over $\delta t$ (Appendix~\ref{sec:fbm_dacf} for further information on DACF).
We measured the DACFs of the simulated AOUP cross-linker ($l_p/\sigma=25$) at various Pe conditions for $\delta t=10^{-2}\tau$ (d) and $10^1\tau $ (e). In the former case where $\delta t<\tau_A$, the negative viscoelastic feedback due to the semiflexible filaments is opposed by the AOUP's directional motion. The viscoelastic feedback is quite strong; the negative dip in the DACF exists even at $\mathrm{Pe}=30$. For $\mathrm{Pe}\geq 60$, the AOUP completely overcomes the negative viscoelastic feedback and its displacements are positively correlated. The profile is reminiscent of the DACFs for actively moving intracellular particles in live amoeba~\cite{Reverey2015}. This effect makes a superdiffusive motion with $\alpha\approx3/2$ at this timescale. In the latter case of $\tau_A<\delta t<\tau_R$ [Fig.~\ref{fig2}(e)], the AOUP cross-linker always experiences strong negative feedback, regardless of Pe, induced by the polymer's viscoelastic response. The negative dip in the DACF gets deeper with a higher Pe, which means that the stronger the AOUP's self-propelled movement the more the anti-persistent displacement of the AOUP. 

Finally, we examined how the active diffusion of the AOUP in semiflexible filaments is changed depending on the number of arms (functionality $f$). 
In Fig.~\ref{fig2}(f) we plot $\alpha$ vs $f$ for the AOUP cross-linkers at $\mathrm{Pe}=30,~60$, and $80$ and the filament stiffness $l_p=25~\sigma$. 
The $\alpha$ tends to increase with the polymer's functionality $f$, which can be understood from Fig.~\ref{fig2}(c) such that the particle feels a stiffer polymer environment as $f$ increases. 
Namely, in the AOUP's viscoelastic diffusion, the relative contribution from the filaments' thermal motion gets stronger than that from the self-propelled motion, which makes the AOUP's active diffusion attain a larger $\alpha$ but with a smaller magnitude of displacements.


\section{Theory: a nonequilibrium fractional Langevin equation }\label{sec:theory}

In this section, we develop a mesoscopic theory for the active diffusion of the AOUP interacting with a semi-flexible filament network observed in Sec.~\ref{sec:result.ABP}. 
In the previous study on the AOUP cross-linker in a flexible star polymer~\cite{joo2020}, we analytically solved the $N$-particle Langevin equation for the AOUP-polymer composite system. It was found that the AOUP's MSD (or velocity autocorrelation) consists of two parts; one from the thermal motion and the other from the self-propelled motion. We found that the observed viscoelastic active diffusion of the AOUP can be described by a fractional Langevin equation with two random noises, Eq.~\eqref{FLEflexible}.   
Here, we apply this idea to the semi-flexible filament system. 
Given that $\mathbf{R}=(X,Y,Z)$  is the Cartesian coordinate of the AOUP interacting with a polymer network, we write down a generalized Langevin equation (GLE) of the form 
\begin{equation}
    \int^t_{-\infty} dt' K(t-t') V(t') = \psi_\mathrm{th}(t) + \psi_\mathrm{ac}(t),
    \label{eq:gle}
\end{equation}
which describes the AOUP's active diffusion under the viscoelastic feedback. Here, we consider the one-dimensional motion; $V=\dot{X}$ is the time derivative of each Cartesian component, e.g., the $x$-component coordinate.
The memory kernel $K(t)$  explains a viscoelastic feedback from the semi-flexible filament network connected to the AOUP cross-linker, which should be determined below. The $\psi_{\mathrm{th}}$ is the thermal noise given to the AOUP at the level of our one-particle description whose covariance is given by the fluctuation-dissipation theorem, i.e., $\langle \psi_\mathrm{th}(t)\psi_\mathrm{th}(t')\rangle =k_BTK(|t-t'|)$. The $\psi_{\mathrm{ac}}$ is an active OU noise at the mesoscopic level, which is in our model assumed to be the bare active noise applied to the AOUP. Then it satisfies the covariance $\langle \psi_\mathrm{ac}(t)\psi_\mathrm{ac}(t')\rangle = \frac{\gamma^2v_p^2}{3}e^{-|t-t'|/\tau_A}$. 

Now we construct the memory kernel based on the tension propagation theory of a semiflexible filament and the topology of a given polymer network. Consider a single semi-flexible filament that undulates in a viscous fluid. Its undulating configuration with time is described by the following Langevin equation
\begin{equation}\label{tension}
\gamma_0 \frac{\partial \mathbf{r}_\perp(s,t)}{\partial t}=k_BTl_p \frac{\partial^4 \mathbf{r}_\perp(s,t)}{\partial s^4}+\bm{\xi}(s,t)
\end{equation} 
where $\mathbf{r}_\perp(s,t)$ is the small undulation of the semi-flexible filament perpendicular to the reference axis at time $t$ and the contour length $s=n\sigma$ ($n$: the monomer index; $\sigma$: filament thickness). $\gamma_0=\gamma/\sigma$ is the frictional coefficient of the filament per unit length. $\bm{\xi}$ is the thermal $\delta$-correlated noise defined above. This equation explains how tension induced by a segmental motion (e.g., the AOUP's motion) propagates along the chain~\cite{sakaue2017active}. Given $n(t)$ is the number of the segments at time $t$ affected by the tension produced by the central monomer (acting as the cross-linker)'s thermal motion at $t=0$, it is related to the time via $\sigma^4 n(t)^4\simeq \frac{k_BTl_p}{\gamma_0}t$. 
This relation allows us to find the microscopic time that the tension is transmitted to the neighboring segment as $\tau_0(=\frac{\sigma^4\gamma_0}{k_BTl_p})=\tau\frac{\sigma}{l_p}$.  Using this,  
we find $n(t)\simeq (\frac{t}{\tau_0})^{1/4}$ and the terminal time
\begin{equation}
\tau_R=\frac{\gamma_0\sigma^4N^4}{k_BT l_p}
\label{tauR}
\end{equation}
that the tension propagates to the end segment, which turns out to be the relaxation time of a semi-flexible chain (Sec.~\ref{sec:result.ABP}). 

Within the tension propagation theory, we further introduce an effective frictional coefficient $\gamma^*(t)$ that the central segment will experience with time $t$, which increases as $\gamma^*(t)\simeq\gamma_0\sigma n(t)=\gamma  (\frac{t}{\tau_0})^{1/4}$ for $\tau_0\ll t\ll \tau_R$. The MSD of the central segment then increases as $\langle \Delta \mathbf{R}_A^2(t)\rangle\simeq \frac{k_BT}{\gamma^*(t)}t \propto t^{3/4}$ for $\tau_0\ll t\ll \tau_R$. We require our GLE \eqref{eq:gle} in the absence of $\psi_\mathrm{ac}$ to give the same diffusion dynamics for the motion of the central segment, which makes us conclude $\gamma^*(t)=t K^{(1)}(t)$ and 
$
K^{(1)}(t)\simeq \frac{\gamma}{\tau_0}\left|\frac{t}{\tau_0}\right|^{-3/4}
$
for $\tau_0\ll t\ll\tau_R$.
Given that our semi-flexible star-like polymer is constructed with $f/2$ chains, the memory kernel of the GLE \eqref{eq:gle} is 
\begin{equation}\label{Kt}
K(t)\simeq \frac{f}{2}\frac{\gamma}{\tau_0}\left|\frac{t}{\tau_0}\right|^{-3/4}.
\end{equation}
Inserting Eq.~\eqref{Kt} into GLE \eqref{eq:gle}, we find that the governing GLE is rewritten as a fractional Langevin equation (FLE)
\begin{equation}\label{FLEstiff}
\frac{\Gamma(1/4)}{2}\frac{f\gamma}{\tau_0^{1/4}}\frac{d^{3/4}X}{dt^{3/4}}=\psi_\mathrm{th}(t)+\psi_\mathrm{ac}(t)
\end{equation}
where $\frac{d^{3/4}X}{dt^{3/4}}$ is the Caputo fractional derivative of order $\frac{3}{4}$~\cite{Metzler2014}. The thermal noise $\psi_\mathrm{th}(t)$ has the zero mean and the covariance $\langle \psi_\mathrm{th}(t)\psi_\mathrm{th}(0)\rangle\propto t^{-3/4}$. This indicates that the $\psi_\mathrm{th}(t)$ is a fractional Gaussian noise with the Hurst exponent $H=5/8(>1/2)$, which is a positively correlated noise. The FLE~\eqref{FLEstiff} is a mesoscopic theory applicable for $\tau_0\lesssim t\lesssim \tau_R$, where tension propagation theory is valid. For alternative analytic schemes to derive similar fractional Langevin equations in different thermal viscoelastic systems, refer to Refs.~\onlinecite{Taloni.1,Taloni.2,Taloni.3}.

Now we solve the FLE \eqref{FLEstiff} [or GLE \eqref{eq:gle}] and compare the diffusion property of this model with that of the AOUP cross-linker studied in the previous section. After rewriting the GLE \eqref{eq:gle} in the Laplace space, we find the mean-squared velocity is given by
\begin{equation}
\left\langle\widetilde{V}
(u)^2\right\rangle=\frac{\langle\widetilde{\psi_{\mathrm{th}}}(u)^2\rangle+\langle\widetilde{\psi_{\mathrm{ac}}}(u)^2\rangle}{\widetilde{K}(u)^2}
\label{eq:ltgle}
\end{equation}
where $\widetilde{f}(u)=\mathcal{L}\{f(t)\}=\int_0^\infty e^{-ut} f(t)dt$.
The GLE \eqref{eq:gle} is driven by two stationary noises, so we can apply the Wiener-Khinchin theorem to relate the mean-squared average of the velocity and the two noises to the respective correlation function. For example, the velocity autocorrelation function (VACF), $C_V(t)=\langle V(t)V(0)\rangle$, satisfies the relation $\langle \widetilde{V}(u)^2\rangle=\widetilde{C_V}(u)/u$~~\cite{pottier2003aging}. Similar relations hold for $\psi_\mathrm{th}(t)$ and $\psi_\mathrm{ac}(t)$. Plugging these relations into Eq.~\eqref{eq:ltgle}, we obtain  
\begin{equation}\label{eq:VACF}
\widetilde{C_V}(u)
=\frac{\widetilde{C_{\psi_{\mathrm{th}}}}(u)+\widetilde{C_{\psi_{\mathrm{ac}}}}(u)}{\widetilde{K}(u)^2}\equiv\widetilde{C_V^\mathrm{(th)}}(u)+\widetilde{C_V^\mathrm{(ac)}}(u).
\end{equation}
Here, $\widetilde{C_{\psi_{\mathrm{th}}}}(u)=k_BT\widetilde{K}(u)\simeq \frac{f}{2}\gamma k_BT\Gamma(1/4) (\tau_0 u)^{-1/4}$ and $\widetilde{C_{\psi_{\mathrm{ac}}}}(u)=\frac{\gamma^2v_p^2}{3}\frac{\tau_A}{1+\tau_A u}$ are the autocorrelation function of each noise term. In the last relation, we emphasize that the VACF consists of the thermal and active components, which are defined as $\widetilde{C_V^\mathrm{(th)}}(u)=\widetilde{C_{\psi_{\mathrm{th}}}}(u)/\widetilde{K}^2$ and $\widetilde{C_V^\mathrm{(ac)}}(u)=\widetilde{C_{\psi_{\mathrm{ac}}}}(u)/\widetilde{K}^2$. Once $\widetilde{C_V}(u)$ is obtained, $C_V(t)$ is 
evaluated through the inverse Laplace transform. As a related quantity, we can evaluate the DACF by integrating the VACF according to Eq.~\eqref{dacf}. As the VACF is composed of the thermal and active components, the DACF reads 
\begin{equation}
C_{\Delta X}(t;\delta t)
=C^\mathrm{(th)}_{\Delta X}(t;\delta t)+
C^\mathrm{(ac)}_{\Delta X}(t;\delta t) 
\label{eq:vact2dacf}
\end{equation}
where each term follows the same asymptotic power-law behavior with the VACF for $t\gg \delta t$. 
Lastly, we obtain the expression for MSD from the VACF as
\begin{equation}
\langle \Delta \widetilde{X^2}(u) \rangle =\frac{2}{u^2}\widetilde{C_V}(u)
=\langle \Delta \widetilde{X^2_\mathrm{th}}(u) \rangle+\langle \Delta \widetilde{X^2_\mathrm{ac}}(u) \rangle.
\end{equation}
The MSD is written as the superposition of its thermal and active parts, each of which is evaluated from the respective VACF [Eq.~\eqref{eq:VACF}].  

\begin{figure}
\centering
\includegraphics[width=8cm]{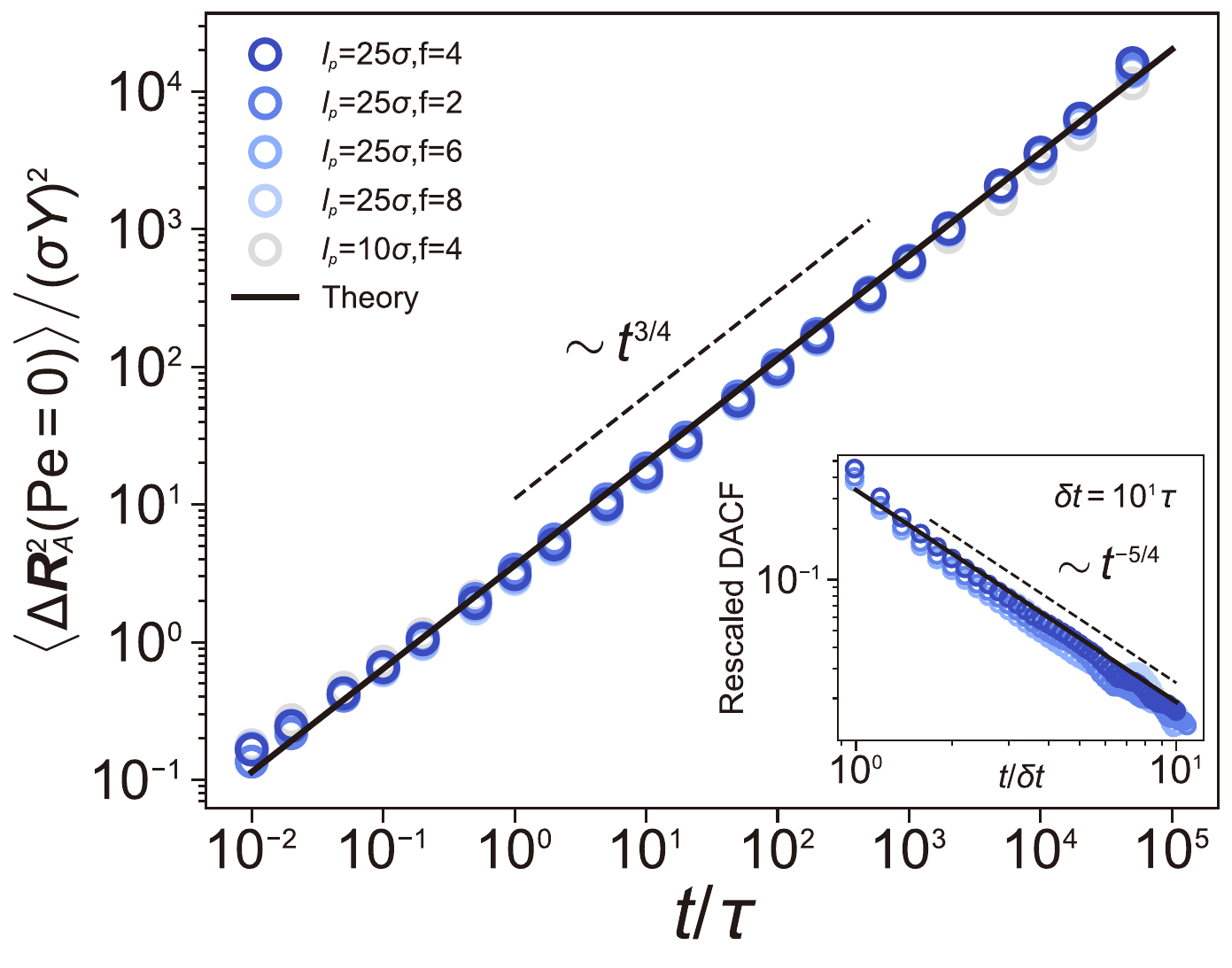}
\caption{
The thermal part of MSD [Eq.~\eqref{eq:msdthermal}] for various simulation conditions ($l_p=10$ and $25$, $f=2,~4,~6$, and $8$). The re-scaling factor is $\sigma^2 Y^2= \frac{\sigma^{9/4}}{l_p^{1/4}f}$. The solid line represents the theoretical curve $\frac{12}{\Gamma(1/4)\Gamma(7/4)} (t/\tau)^{3/4}$.
Inset: The DACFs [Eq.~\eqref{eq:dacf_th}] for the five cases in (a) at $\delta t=10\tau$. The $y$-axis is plotted with 
$\mathrm{DACF}/[-(\sigma Y)^2 (\delta t/\tau)^{3/4}]$, which is collapsed on the theoretical curve (solid) $-\frac{6}{\Gamma(1/4)\Gamma(-1/4)} (t/\delta t)^{-5/4}$. 
The dashed line is a guide for the expected scaling.
}
\label{fig3}
\end{figure}

\textit{Thermal part.---} 
In the absence of the active noise, the VACF is solely determined by the thermal part, 
$
\widetilde{C_V}^\mathrm{(th)}(u)
=\frac{k_BT}{\widetilde{K}(u)},
$
which yields
\begin{equation}
C^\mathrm{(th)}_{V}(t)\simeq \frac{2}{\Gamma(1/4)\Gamma(-1/4)} 
(\sigma Y)^2 \tau^{-2} 
\left(\frac{t}{\tau}\right)^{-5/4}
\label{eq:vacf_th}
\end{equation}
for $\tau_0\ll t\ll\tau_R$ where $Y=\left(\frac{\sigma}{l_pf^4}\right)^{1/8}$.
Here, 
$\Gamma(-1/4)=-\frac{16}{3}\Gamma(7/4)<0$ 
indicates that the thermal viscoelastic motion of the cross-linker is anti-persistent. Note that the amplitude of $C^\mathrm{(th)}_{V}(t)$ decays as $1/[f l_p^{1/4}]$.
Using Eq.~\eqref{eq:vacf_th} we obtain the DACF
\begin{equation}
\begin{split}
C_{\Delta X}^{\mathrm{(th)}}(t;\delta t)
    \simeq
    \frac{2}{\Gamma(1/4)\Gamma(-1/4)} 
    (\sigma Y)^2 (\delta t/\tau)^{3/4}
    \left(\frac{t}{\delta t}\right)^{-5/4}
\end{split}
\label{eq:dacf_th}
\end{equation}
for $\delta t\ll t\ll \tau_R$.
In Fig.~\ref{fig3}(inset), we plot the simulated DACFs for various simulation conditions with $\delta t=10\tau$. To confirm Eq.~\eqref{eq:dacf_th}, we rescale the DACFs with the factor of 
$(\sigma Y)^2 (\delta t/\tau)^{3/4}$ and observe the collapse of the simulation data onto the theoretical curve 
$-\frac{6}{\Gamma(1/4)\Gamma(-1/4)} (t/\delta t)^{-5/4}$.

The thermal part of the MSD grows as
\begin{equation}
\langle\Delta X_\mathrm{th}^2(t)\rangle
\simeq
\frac{4}{\Gamma(1/4)\Gamma(7/4)} 
(\sigma Y)^2
\left(\frac{t}{\tau}\right)^{3/4}.
\label{eq:msdthermal}
\end{equation}
The power-law exponent $3/4$ is the well-known exponent for undulations of a semi-flexible chain. In Fig.~\ref{fig3}(a) we plot the rescaled MSD of Brownian cross-linkers at various $f$ and $l_p$. They all collapse on the master curve (solid line), $\frac{12}{\Gamma(1/4)\Gamma(7/4)} (t/\tau)^{3/4}$, on the timescales of $\tau_0\ll t \ll \tau_R$. 
Note that the thermal motion has the same MSD and DACF with fractional Brownian motion with $H=3/8$ (Appendix~\ref{sec:fbm_dacf}). 

\begin{figure}
    \centering
    \includegraphics[width=8cm]{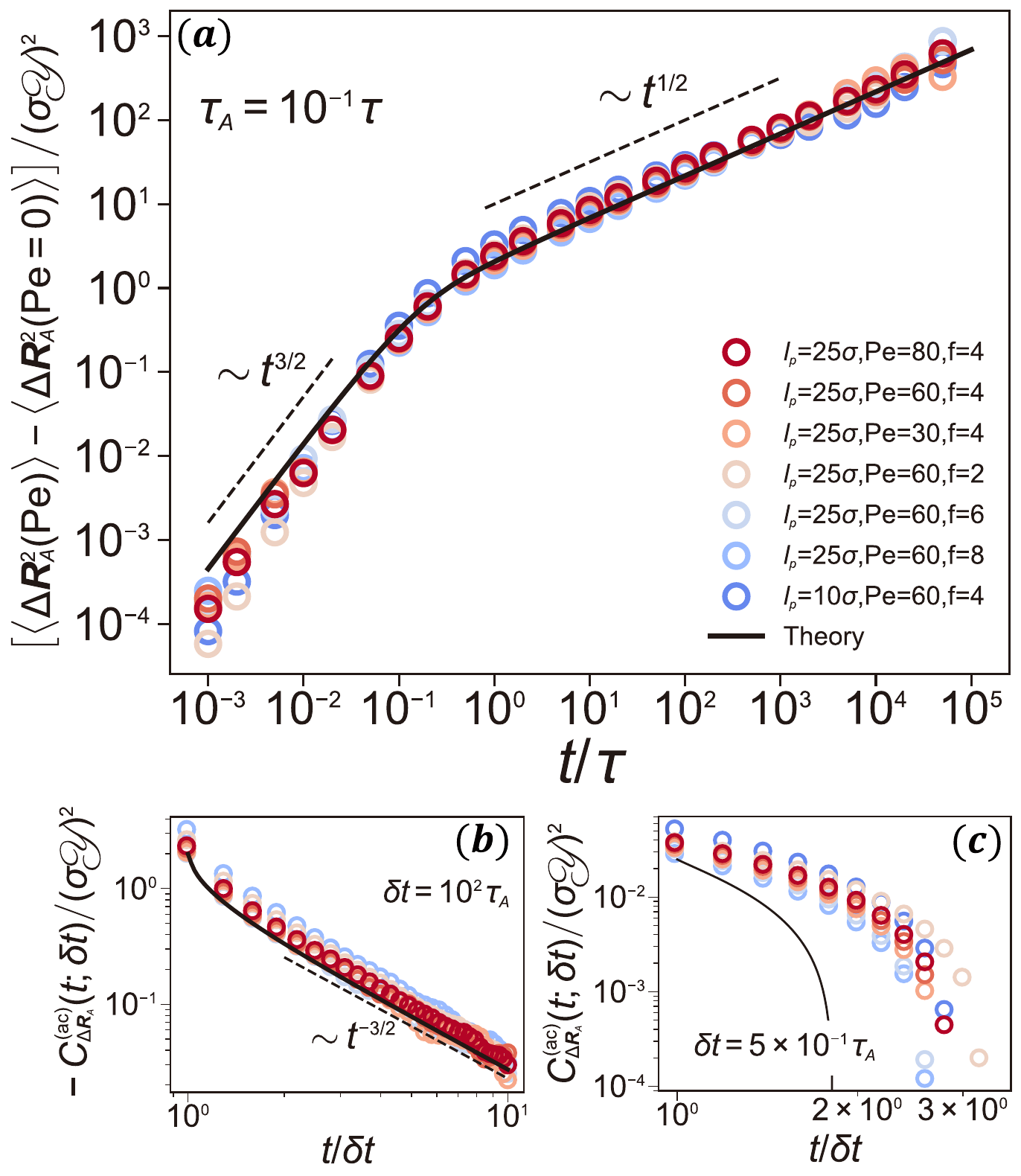}
    \caption{
    (a)
    The active part of MSD, $\langle \Delta \mathbf{R}_A^2 (t;\mathrm{Pe})\rangle-\langle \Delta \mathbf{R}_A^2 (t;\mathrm{Pe}=0)\rangle$, for various simulation conditions ($l_p=10,~25$, $\mathrm{Pe}=30,~60,~80$, $f=2,~4,~6$, and $8$). They are rescaled with the factor of $\sigma^2 \mathcal{Y}^2$ [see Eq.~\eqref{Y}].  
    The solid line is the theoretical expectation,  $\frac{16}{\Gamma^2(1/4){\pi}^{1/2}}[\sqrt{t/\tau_A}-D_+(\sqrt{t/\tau_A})]$, from Eq.~\eqref{eq:msdactive}.
    (b) \& (c)
    The active parts of the DACF. The symbols are from the same simulation condition in (a). The time lags are (b) $\delta t=10^{2}\tau_A$ and (c) $\delta t=5\times 10^{-1}\tau_A$. The solid line shows theory [Eq.~\eqref{eq:dacf_ac}].
    }
    \label{fig4}
\end{figure}

\textit{Active part.---} The active part of the VACF is
$
\widetilde{C_V}^\mathrm{(ac)}(u)=4v_p^2\tau_A/[3f^2\sigma^2\Gamma^2(1/4)(1+\tau_A u)(u\tau_0)^{-1/2}].
$
By the inverse Laplace transform, we find
\begin{equation}
\begin{split}
C_{V}^{\mathrm{(ac)}}(t)&\simeq 
\frac{8}{3\Gamma^2(1/4){\pi}^{1/2}}
\frac{\tau_A^{-1/2} }{{\tau}^{3/2}} 
\frac{\sigma^{5/2}}{l_p^{1/2}}
\frac{\mathrm{Pe}^2}{f^2}
\\&\qquad\qquad
\times\left(
\frac{1}{2}\frac{1}{\sqrt{ t/\tau_A}}-
D_+\left(\sqrt{t/\tau_A}\right)
\right)\\
&\simeq
\frac{8}{3\Gamma^2(1/4){\pi}^{1/2}}
{(\sigma\mathcal{Y})}^2\tau_A^{-2}
\\&\qquad\qquad
\times\left\{\begin{array}{lr}
\frac{1}{2}
\left(\frac{t}{\tau_A}\right)^{-1/2},  
& \tau_0\ll t\ll\tau_A\\
\Bigl(-\frac{1}{4}\Bigr)
\left( \frac{t}{\tau_A}\right)^{-3/2},
& \tau_A \ll t\ll\tau_R
\end{array}\right.
\end{split}
\label{eq:vacf_ac}
\end{equation}
where $D_+(z)= \frac{1}{2}\int^\infty_0 e^{-t^2/4}\sin(zt)dt$ 
is the Dawson function, the one-sided Fourier-Laplace sine transform of a Gaussian function.
Note that for simplicity we introduce a dimensionless constant
\begin{equation}\label{Y}
\mathcal{Y}
=\biggl(\frac{\tau_A}{\tau}\biggr)^{3/4}\biggl(\frac{\sigma}{l_p}\biggr)^{1/4}\frac{\mathrm{Pe}}{f},
\end{equation}
which plays a role as a scaling factor in physical observables presented below. 
Notably, the active VACF has two distinct power-law scalings shown in the second line. For $t\ll \tau_A$, $D_+(\sqrt{t/\tau_A})\approx (t/\tau_A)^{1/2}-\frac{2}{3}(t/\tau_A)^{3/2}\ll 1/\sqrt{t/\tau_A}$. Thus, the active part of VACF decays as $t^{-1/2}$ at short times with the positive sign, indicating that the active viscoelastic motion is persistent. When $t\gg \tau_A$,  $D_+(\sqrt{t/\tau_A})\approx \frac{1}{2}(t/\tau_A)^{-1/2}+\frac{1}{4}(t/\tau_A)^{-3/2}$, so the VACF decays as $-t^{-3/2}$. This means that after the propulsion time, $\tau_A$, the viscoelastic feedback leads to anti-persistent movement with a power-law exponent $3/2$ (which is different from the thermal case $5/4$).
Similarly, we obtain the expression for DACF
\begin{equation}
\begin{split}
    C_{\Delta X}^{\mathrm{(ac)}}(t;\delta t)
    &\simeq 
         \frac{8}{3\Gamma^2(1/4){\pi}^{1/2}}
         (\sigma\mathcal{Y})^2
         \\&\qquad\quad
        \times\left[
            M\left(\frac{|t-\delta t|}{\tau_A}\right)
            +M\left(\frac{t+\delta t}{\tau_A}\right)
            -2M\left(\frac{t}{\tau_A}\right)
        \right]
    \\&\simeq
    \frac{8}{3\Gamma^2(1/4){\pi}^{1/2}}
    (\sigma\mathcal{Y})^2(\delta t/\tau_A)^2
    \\&\qquad\quad
    \times\left\{\begin{array}{lr}
    \frac{1}{2}\left( \frac{t}{\tau_A}\right)^{-1/2},
    & \delta t\ll t \ll \tau_A\\
    \Bigl(-\frac{1}{4}\Bigr)\left(\frac{t}{\tau_A}\right)^{-3/2},  
    & \tau_A \ll \delta t\ll t
    \end{array}\right.
\end{split}
\label{eq:dacf_ac}
\end{equation}
where $M(z)=\sqrt{{z}}-D_+\left({\sqrt{z}}\right)$ is used.
The DACF has the same power-law scaling relation as the VACF with a different amplitude. To examine whether the simulation data for various parameter conditions follow Eq.~\eqref{eq:dacf_ac}, we plot the simulated DACFs, $C_{\Delta X}^{\mathrm{(ac)}}(t;\delta t)/(\sigma\mathcal{Y})^2$, to collapse on the master curve (solid line) given by Eq.~\eqref{eq:dacf_ac} [Fig.~\ref{fig4}(b) \& (c)].   
The results show the following. 
For the timescales of $\delta t > \tau_A$ [Fig.~\ref{fig4}(b)], the plotted DACFs have the same power-law decay ($t^{-3/2}$) and are in excellent agreement with the theory (solid line). 
However, for the timescales of $\tau_0<\delta t<\tau_A$ [Fig~\ref{fig4}(c)], the DACFs do not precisely follow the expected scaling (solid line). The FLE model does not perfectly fit in this regime, presumably because the condition of $\delta t\gg \tau_0$ was insufficiently met where the active ballistic motion is not fully transmitted along the polymer particles, so the viscoelastic feedback is incomplete.

From the VACF \eqref{eq:vacf_ac} we evaluate the MSD to find 
\begin{equation}
\begin{split}
    \langle \Delta X_{\mathrm{ac}}^2(t)\rangle 
    &\simeq 
         \frac{16}{3\Gamma^2(1/4){\pi}^{1/2}}
         (\sigma\mathcal{Y})^2
         M\left(
            \sqrt{t/\tau_A}
         \right)\\
    &\simeq
    \frac{16}{3\Gamma^2(1/4){\pi}^{1/2}}
    (\sigma\mathcal{Y})^2
    \\&\qquad\qquad\qquad
    \times\left\{\begin{array}{lr}
    \frac{2}{3}\left(\frac{t}{\tau_A}\right)^{3/2},  
    & \tau_0 \ll t\ll\tau_A\\
    \left( \frac{t}{\tau_A}\right)^{1/2},
    & \tau_A\ll t \ll \tau_R
    \end{array}\right.
\end{split}
\label{eq:msdactive}
\end{equation}
The MSD is expected to increase as $\sim t^{3/2}$ for $\tau_0<t<\tau_A$ and $\sim t^{1/2}$ for $t>\tau_A$. 
Interestingly, the active displacement also behaves as fractional Brownian motion in terms of the power-law scaling for the MSD and DACF ($H=3/4$ for $\tau_0<t<\tau_A$ and $H=1/4$ for $t>\tau_A$).
In Fig.~\ref{fig4}(a) we plot the MSDs for various simulation conditions after rescaling with 
$\sigma^2\mathcal{Y}^2$. 
The simulation data overall follow the scaling behaviors with the expected amplitude. We note that if Pe is sufficiently high in which the active MSD dominates over the passive counterpart, the active particle interacting with a semi-flexible chain seemingly illustrates the Rouse motion ($t^{1/2}$) of a flexible chain for $t>\tau_A$. Here, the active Rouse-like diffusion is distinguished from the genuine Rouse motion via the magnitude of displacements. The fact that the active exponent $1/2$ is smaller than the thermal part $3/4$ suggests that in the semi-flexible polymer media the active displacement is more subdiffusive than the thermal displacement. This behavior is seen as counter-intuitive in that the persistent active noise helps the particle diffuse faster than the thermal particle fluctuating by a $\delta$-correlated noise. We also note that the superdiffusion for $t<\tau_A$ occurs with the anomalous exponent slightly greater than $3/2$. This discrepancy is also evident in our analysis of the DACF at these timescales [Fig.~\ref{fig4}(c)], which presumably stems from inaccuracies in the self-propulsion time or noise strength of the active noise that can play a significant role in the displacement correlation at these specific timescales.

\begin{figure}
\centering
\includegraphics[width=8cm]{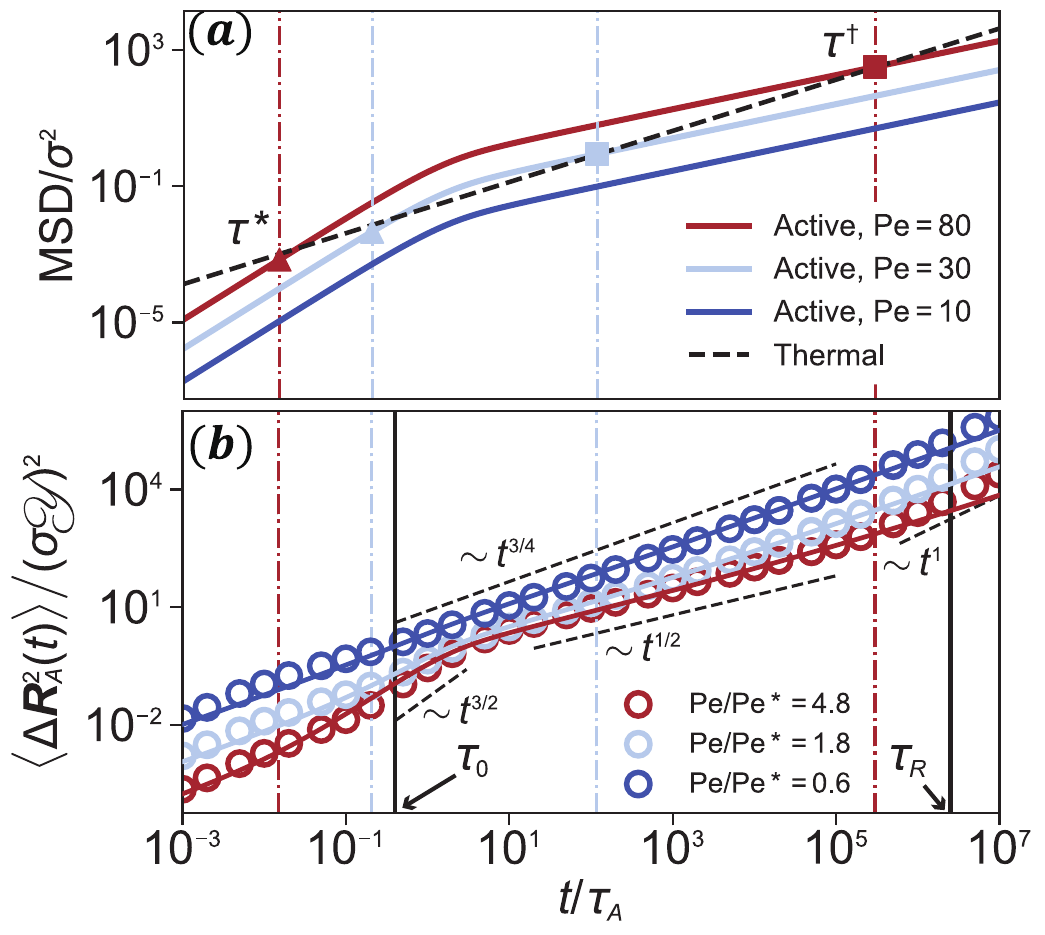}
\caption{The comparison of the thermal and active  displacements. (a) Theoretical MSD curves for the thermal (dashed, Eq.~\eqref{eq:msdthermal}) and active (solid, Eq.~\eqref{eq:msdactive}) motion for $\mathrm{Pe}=10,~30$, and 80. Two cross-over times, $\tau^*$ (triangle) and $\tau^\dagger$ (square), are annotated. (b) The simulated (circle) and theoretical (solid line) MSDs for the AOUP cross-linker at the same Pe condition in (a). Here, Pe is annotated in terms of Pe$^*$.}
\label{fig5}
\end{figure}

Now let us combine the thermal and active contributions and investigate the superimposed motion, which represents the diffusion dynamics of the AOUP cross-linker in $[\tau_0,~\tau_R]$. The active dynamics is manifested only for sufficiently high Pe conditions and at specific timescales. We illustrate this idea in Fig.~\ref{fig5}(a) where the theoretical MSD curves for the thermal (dashed line) [Eq.~\eqref{eq:msdthermal}] and active (solid lines) [Eq.~\eqref{eq:msdactive}] components are separately plotted. The plot shows that when $\mathrm{Pe}=10$ the thermal displacements are always larger than the active counterpart. Their sum, i.e., the MSD of the AOUP cross-linker, is thus dominated by the thermal motion, which is confirmed in the simulated total MSD at this condition plotted in Fig.~\ref{fig5}(b) [$\mathrm{Pe}/\mathrm{Pe}^*=0.6$]. Here, the AOUP cross-linker displays seemingly the MSD ($\sim t^{3/4}$) of a thermal particle in a semiflexible filament as the active displacement is neglected compared to the thermal motion.
For $\mathrm{Pe}=30$ and $80$ [Fig.~\ref{fig5}(a)], the thermal and active MSDs cross at two time points, and we define them as the cross-over time $\tau^*(<\tau_A)$ and $\tau^\dagger(>\tau_A)$. It is important to note that the active dynamics is only dominant in between $\tau^*$ and $\tau^\dagger$. 
The two cross-over times are found by solving $\widetilde{C_{\psi_{\mathrm{th}}}}(u) =\widetilde{C_{\psi_{\mathrm{ac}}}}(u)$, which yields 
\begin{equation}\label{tau*}
   \tau^*/\tau_A
=
\left(
\frac{2}{3\Gamma(1/4)}
\mathrm{Pe}\ \mathcal{Y}
\right)^{-4/3} 
\end{equation}
and 
\begin{equation}\label{taudag}
    \tau^\dagger/\tau_A
=
\left(\frac{2}{3\Gamma(1/4)}
\mathrm{Pe} \mathcal{Y}
\right)^{4}. 
\end{equation}
At sufficiently high Pe conditions, the two cross-over times are well separated in which the active dynamics are observed in between them. In Fig.~\ref{fig5}(b), we crosscheck the validity of the FLE by comparing our theory to the simulation data at three distinct Pe values. Here, the solid lines are the analytic expression [the sum of Eqs.~\eqref{eq:msdthermal} and \eqref{eq:msdactive}] for $\langle \mathbf{R}_A^2(t) \rangle$. For $\tau_0\lesssim t\lesssim \tau_R$, the theory gives good agreement with the simulation. See also $\tau^*$ and $\tau^\dagger$, annotated with the dash-dotted lines in the plot, for $\mathrm{Pe}/\mathrm{Pe}^*=1.8$ and 4.8. In accordance with the above theory, the active dynamics is visible within the window of $[\tau^*,~\tau^\dagger]$, where the cross-over times sensitively depend on Pe. 

As Pe is decreased to the threshold value $\mathrm{Pe}^*$ satisfying the condition 
\begin{equation}\label{Pe*}
\frac{2}{3\Gamma(1/4)}\mathrm{Pe}^*\mathcal{Y}=1,
\end{equation} 
the two cross-over times are equal to each other ($\tau^*=\tau^\dagger=\tau_A$), and the active motion becomes negligible compared to the thermal motion. Therefore, the $\mathrm{Pe}^*$ is considered as the threshold Pe above which the active dynamics start to dominate over the thermal part. For the given simulation parameters used  in Fig.~\ref{fig5}(b), $\mathrm{Pe}^*$ is estimated to be about $17$, consistent with the simulation results. We also note that even though $\tau^\dagger(\sim \mathrm{Pe}^8)$ rapidly increases with $\mathrm{Pe}$, the active Rouse motion ($t^{1/2}$) is manifested only until the maximum tension propagation time (the relaxation time) $\tau_R$, and the active cross-linker shows a Fickian motion for $t>\tau_R$. 

\section{Concluding remarks} \label{sec:summary}
In this work, we have investigated the active diffusion performed by an AOUP cross-linking (or strongly bound to) a semi-flexible star-like polymer by means of the Langevin dynamics simulation and the FLE \eqref{FLEstiff}.
When the functionality is $f=2$, our model describes the undulation dynamics of a single semi-flexible filament where the center monomer is the AOUP. We observed that the AOUP connected to a stiff filament exhibits a superdiffusion of anomalous exponent $\alpha\approx 3/2$ for $t<\tau_A$ and a subdiffusion with $1/2\leq\alpha\leq 3/4$ for $t>\tau_A$.

Analogously to the AOUP cross-linker in a flexible polymer system~\cite{joo2020}, the AOUP in the semi-flexible environment becomes more subdiffusive for $t>\tau_A$ as it diffuses with a higher Pe. An important finding was that the exponent converges to the Rouse exponent $1/2$ of a thermal flexible chain as $\mathrm{Pe}\gg1$. This result was consistently observed regardless of the functionality and the boundary condition. This finding may be relevant to interpreting \textit{in vivo} filament or transport dynamics observed in experiments~\cite{Stadler2017,Speckner2018}.

We demonstrated that the nonequilibrium viscoelastic diffusion of an AOUP embedded in a semi-flexible filament network can be described by the FLE~\eqref{FLEstiff}. It is a generalized Langevin equation with a memory kernel $K(t)\propto t^{-3/4}$ explaining the viscoelastic feedback from the semi-flexible filaments, combined with a fractional Gaussian thermal noise of the Hurst exponent $H=5/8$ and an Ornstein-Uhlenbeck active noise. It turns out that while the thermal noise always induces an anti-persistent viscoelastic motion, the active noise results in a time-varying viscoelastic motion. It is a persistent superdiffusive motion of $\alpha=3/2$ for $t\lesssim \tau_A$ and an anti-persistent subdiffusive motion of $\alpha=1/2$ for $t\gtrsim \tau_A$. Due to the thermal motion, the active viscoelastic dynamics is not always seen. It is only visible for Pe values larger than the threshold one $\mathrm{Pe}^*$ [Eq.~\eqref{Pe*}] and within the two cross-over times $\tau^*$ and $\tau^\dagger$ [Eqs.~\eqref{tau*} and \eqref{taudag}]. 

There are a few comments on the scope of our model and theory.
While our study is based on the AOUP model, the observed collective active dynamics may occur in other active particle models. This view is corroborated by our supplementary simulation study of the active cross-linker using the active Brownian particle (ABP) model. The MSDs of the AOUP cross-linker simulated in Fig.~\ref{fig2}(a) can be reproduced with the ABP cross-linker, see Fig.~\ref{figa1}(e). The AOUP in the model system may be either a self-propelling entity or a nonequilibrium correlated noise from external sources. In the former case, the model describes a Janus particle (or a motile cell) strongly attached to a part of a polymer network or stuck in a concentrated polymer gel. In the latter case, it could represent active forces generated by motor proteins in a cell or even an active bath itself. As noted in Ref.~\onlinecite{Woillez2020}, the net force of randomly oriented pulses having a finite duration time exerted by multiple motor proteins could be approximated by an active Ornstein-Ulhenbeck noise. Indeed, it was reported in a seminal study on the active intracellular transport~\cite{Caspi2000} that the actively transported tracer in a cell exhibits the active viscoelastic diffusion described in our model, i.e., a superdiffusion of $\alpha=3/2$ followed by a subdiffusion of $\alpha=1/2$. Our study strongly suggests that the observed dynamics was, in fact, a confined active motion of a microbead driven by motor proteins before escaping from a local trap associated with semi-flexible filaments.

Beyond the scope of the current study, we expect that the GLE description can be expanded to account for active particles in more complicated polymer networks than the current model or in intracellular environments that may include the effect of macromolecular crowding. These complicated effects change the viscoelastic response of the system, which may be effectively captured by the properties of the memory kernel $K(t)$ in the GLE formalism. It is an open question to determine the appropriate memory kernel for a given viscoelastic system. Recent work has shown that for a relatively simple elastic chain that can stretch and bend, the $K(t)$ can be mathematically derived with two distinct power-law regimes arising from the bending and stretching responses at earlier and later times, respectively \footnote{A detailed report of this study will be published elsewhere}. 
For more complicated viscoelastic systems, the $K(t)$ many need to be inferred empirically from a microrheology study of the environment. Future work will focus on testing and extending the current theoretical framework to a variety of active viscoelastic systems.

\section*{Acknowledgments}
This work was supported by the National Research Foundation
(NRF) of Korea, Grant RS-2023-00218927, and JSPS KAKANHI (Grants No.~JP18H05529 and JP21H05759). We thank
Xavier Durang for valuable comments on the manuscript.

\section*{Author declarations}
\subsection*{Conflicts of Interest }
There are no conflicts to declare.  



\appendix
\renewcommand\thefigure{A\arabic{figure}}
\newcommand{\theHfigure}{A\arabic{figure}}
\setcounter{figure}{0} 


\begin{figure}
    \centering
    \includegraphics[width=8cm]{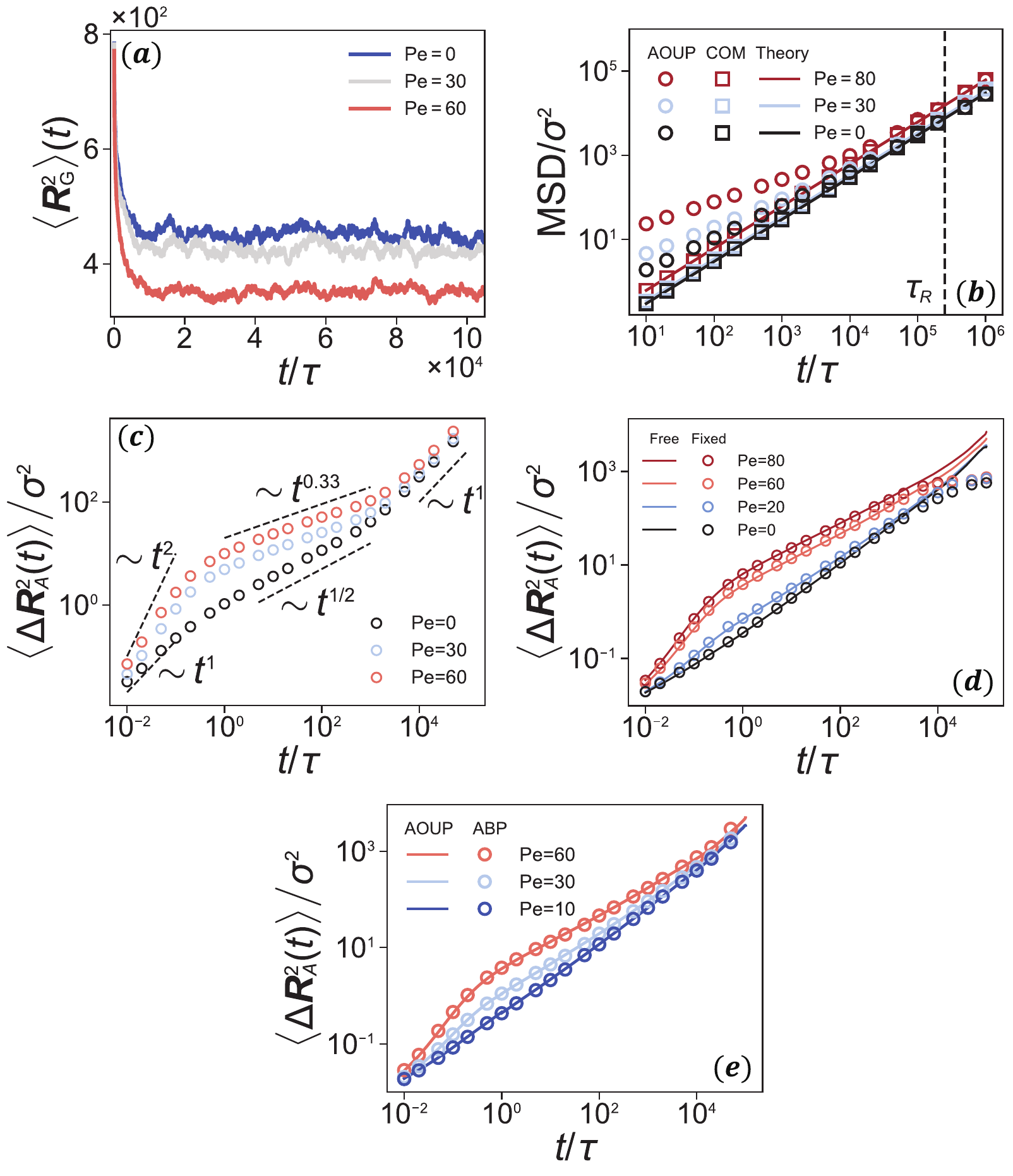}
\caption{
(a) The simulated mean-squared radius of gyration of the active star polymer $\textbf{R}^2_G (t)$ for three Pe values. At $t=0$, the polymer arms were fully stretched. The system reaches the stationary state after $t\approx 1.6\times 10^4 \tau$.
(b) The simulated MSDs for the AOUP cross-linker (circle) and the center-of-mass (square) of the active star polymer. It can be inferred that the Fickian dynamics of the AOUP for $t\gg \tau_R$ is attributed to the drift of the center-of-mass (COM). The solid lines depict the theoretically expected diffusivity of the COM, $6D_G+2v_p^2\tau_A/N_\mathrm{tot}^2$. 
(c) The MSD curves for the AOUP cross-linker for $l_p/\sigma=0.5$. 
(d) The comparison of the MSD curves for the AOUP cross-linker with the free B.C. (line) and fixed B.C. (symbol). The same semiflexible filament system as in Fig.~2 ($f=4$ and $l_p/\sigma=25$). 
(e) The comparison of the simulated MSDs for the AOUP cross-linker (solid line) and the active Brownian particle (ABP) cross-linker (circle). ABPs were simulated via the self-propulsion movement and rotational diffusion of the particle introduced in Ref.~\onlinecite{Zheng2013}. The ABP’s propulsion memory time was set to the same as that of the AOUP by controlling the rotational diffusivity of the ABP, $D_r=1/(2\tau_A)$
}
\label{figa1}
\end{figure}


\section{The AOUP cross-linker in the flexible chain limit} \label{sec:flex}


In Fig.~\ref{figa1}(c), the MSDs in the flexible chain limit ($l_p/\sigma=0.5$) are plotted for various Pe values. The dashed lines show the scaling guide for anomalous exponents. We can observe the well-known thermal Rouse dynamics ($\alpha=1/2$) at $\mathrm{Pe}=0$. At the same time scale, the $\alpha$ monotonically decreases as $\mathrm{Pe}$ increases. The guideline shows $\alpha\approx0.33$ at $\mathrm{Pe}=60$, which is less than the thermal value. Here we reproduced the active subdiffusion for the AOUP cross-linker in a flexible star-like polymer reported in Ref.~\onlinecite{joo2020}.
At short times shorter than $\tau_A$, the AOUP shows a super-diffusive motion of $\alpha\approx 2$ while the Brownian cross-linker has $\alpha\approx1$. 
After the Rouse relaxation time, the AOUP exhibits Fickian dynamics. 

\section{The simulation result with the fixed boundary condition}\label{sec:boundary}

To see the boundary effect on the AOUP dynamics, we repeated the simulation with the fixed boundary condition. Figure~\ref{figa1}(d) shows the comparison of the MSDs for the AOUP cross-linker for the two B.C.s (i.e., the fixed and free boundary conditions).  
The boundary conditions were irrelevant in that the two cases display identical AOUP dynamics up to  $t\sim \tau_R$. The boundary effect is only trivially visible for $t\gg \tau_R$ where the particle exhibited the Fickian diffusion for the free B.C. and the confined diffusion for the pinned end monomer. 
This is because, prior to the tension from the AOUP propagating to the end of the polymer, the neighboring monomers that interact with the AOUP exhibit the same collective motion regardless of the polymer's end state. Thus, the FLE~\eqref{FLEstiff} governs both systems, and the boundary condition is irrelevant to the viscoelastic active diffusion observed in the main text.

\section{Displacement autocorrelation for fractional Brownian motion}\label{sec:fbm_dacf} 
Fractional Brownian motion (FBM) is a stationary-incremental but correlated Gaussian process $X_H(t)$ characterized by the autocorrelation \cite{FBM1968,Burov2011,Molz1997} 
\begin{equation}\label{eq:fbm}
\langle X_H(t)X_H(t')\rangle\propto (|t|^{2H}+|t'|^{2H}-|t-t'|^{2H})
\end{equation}
where $H$ is referred to as the Hurst exponent in $(0,~1)$.
The mean-squared displacement increases as $\langle X_H^2(t)\rangle=2D_H t^{2H}$ where $D_H$ is the generalized diffusivity of physical dimension $[\mathrm{m}^2/\mathrm{s}^{2H}]$. FBM is subdiffusive for $H<1/2$, Fickian at $H=1/2$, and superdiffusive for $H>1/2$. 

The increment $\xi_H(t)=dX_H(t)/dt$ is known as fractional Gaussian noise (FGN). The velocity autocorrelation of FBM is then the autocorrelation of FGN, i.e., 
\begin{equation}\label{eq:FGN}
    \begin{split}
         \langle \xi_H(t)\xi_H(t')\rangle
         =& 2D_{2H}H(2H-1)|t-t'|^{2H-2}
         \\
         &+4D_{2H}H|t-t'|^{2H-1}\delta(t-t').
    \end{split}
\end{equation}
For normal diffusion at $H=1/2$, $\langle \xi_H(t)\xi_H(t')\rangle=2D_{1}\delta(t-t')$. For $H\neq 1/2$, $\langle \xi_H(t)\xi_H(t')\rangle\sim (2H-1)|t-t'|^{2H-2}$. Note that, except for the Brownian motion at $H=1/2$, FBM has a power-law decaying velocity autocorrelation. The prefactor indicates that FBM is anti-persistent for $0<H<1/2$ and persistent for $1/2<H<1$.  

For a given time lag $\delta t$, we define a displacement $\Delta X(t;\delta t)=X_H(t+\delta t)-X_H(t)$. Its autocorrelation, i.e., the displacement autocorrelation (DACF) for FBM, is evaluated from Eq.~\eqref{eq:fbm}. The normalized DACF reads \cite{Jeon2012}
\begin{equation}\label{eq:DACF}
\begin{split}
   \frac{C_{\Delta X}(t;\delta t)}{C_{\Delta X}(0;\delta t)} &=\frac{(t+\delta t)^{2H}-2t^{2H}+|t-\delta t|^{2H}}{2\delta t^{2H}}
   \\
   & \propto {(2H-1)}({t}/{\delta t})^{2H-2}~\hbox{for}~t/\delta t\gg 1.
\end{split}
\end{equation}
Note that in the large-time limit the DACF has the same correlation structure with the velocity autocorrelation function above. For $0<H<1/2$, the autocorrelation is negative, indicating that any two displacements separated by $t$ are likely in the opposite direction. For $1/2<H<1$, the autocorrelation is positive and any two displacements of FBM tend to be in the same direction.   


\bibliographystyle{apsrev4-1} 
\bibliography{./arxiv_revision.bbl}
\end{document}